\documentclass[12pt]{article}

\usepackage{amssymb}
\usepackage{epsfig}

\newcommand{\re}{\mathrm{Re}\,}
\newcommand{\im}{\mathrm{Im}\,}

\newcommand{\gev}{\,{\rm GeV}}

\begin{document}

\begin{flushright}
DESY 04-238 \\
hep-ph/0412162 \\
\end{flushright}

\begin{center}
\vskip 2.5\baselineskip
{\LARGE \bf
Quarks vs.\ gluons \\[0.5em]
in exclusive $\rho$ electroproduction
}

\vskip 2.5\baselineskip
M.~Diehl \\[0.5\baselineskip]
\textit{Deutsches Elektronen-Synchroton DESY, 22603 Hamburg, Germany}
\\[1.5\baselineskip] 
A.~V.~Vinnikov \\[0.5\baselineskip]
\textit{Bogoliubov Laboratory of Theoretical Physics, JINR, 141980
  Dubna, Russia \\ and \\
Deutsches Elektronen-Synchroton DESY, 15738 Zeuthen, Germany}
\vskip 3\baselineskip
\textbf{Abstract}\\[0.5\baselineskip]
\parbox{0.9\textwidth}{\small We compare the contributions from quark
and from gluon exchange to the exclusive process $\gamma^* p\to \rho^0
p$.  We present evidence that the gluon contribution
is substantial for values of the Bjorken variable $x_B$ around 0.1.
\vskip 2\baselineskip
}
\vskip 1.5\baselineskip
\end{center}

%%%%%%%%%%%%%%%%%%%%%%%%%%%%%%%%%%%%%%%%%%%%
\noindent 1.\,
There is an ongoing experimental and theoretical effort to determine
generalized parton distributions \cite{mul:94,ji:96} from hard
exclusive processes like deeply virtual Compton scattering and
electroproduction of mesons.  These distributions encode fundamental
information about nucleon structure, in particular about the angular
momentum carried by partons \cite{ji:96} and about their spatial
distribution \cite{ral:02,bur:02}.  An important process is the
production of $\rho^0$ mesons, well suited for experimental study
because of its relatively high cross section and its clean final state
signature from the decay $\rho^0\to \pi^+\pi^-$.  As pointed out in
\cite{goe:01}, the transverse target polarization asymmetry of this
channel is sensitive to the nucleon spin-flip distribution $E$
appearing in the angular momentum sum rule \cite{ji:96}.

\begin{figure}
\begin{center}
\includegraphics[width=0.8\textwidth]{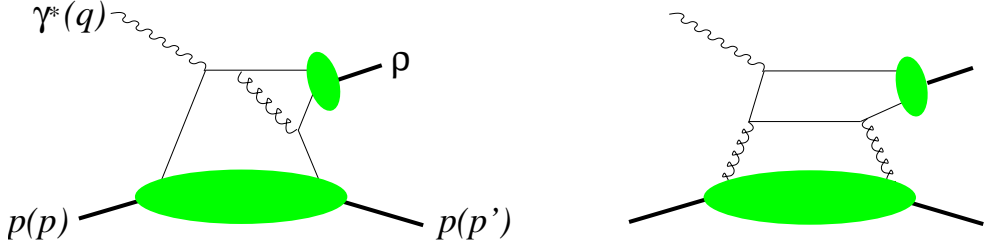}
\caption{\label{fig:mesons} Example graphs for $\gamma^* p \to \rho\,
p$ with generalized quark and gluon distributions.  Four-momenta are
given in parentheses.}
\end{center}
\end{figure}

Quark and gluon distributions contribute to $\rho$ production at the
same order in $\alpha_s$, as seen in Fig.~\ref{fig:mesons}.  For the
separation of quark and gluon degrees of freedom this channel is thus
a valuable complement to deeply virtual Compton scattering, which
offers the cleanest and most detailed access to generalized parton
distributions \cite{Belitsky:2001ns,die:03}, but is sensitive to
gluons only at the level of $\alpha_s$ corrections.  {}From the
behavior of the usual quark and gluon densities one expects $\rho$
production to be dominated by gluons at very small $x_B$ and by quarks
at very large $x_B$, and it is natural to ask where the transition
between these two regimes takes place.  In this letter we present
evidence that quarks and gluons contribute to the $\rho$ cross section
with comparable strength in the $x_B$ region around 0.1, relevant for
measurements at HERMES \cite{hermes:00}.  Key ingredient in our
argument is the measured cross section for $\phi$ electroproduction,
where the gluon distribution should dominate.

%%%%%%%%%%%%%%%%%%%%%%%%%%%%%%%%%%%%%%%%%%%%
\vskip \baselineskip
\noindent 2.\,
We consider the exclusive processes $\gamma^* p\to \rho\, p$ and
$\gamma^* p \to \phi\, p$ and use the standard kinematic variables
$Q^2= -q^2$, $W^2= (p+q)^2$, $x_B = Q^2 /(2 p\cdot q)$ and
$t=(p-p')^2$.  In the limit of large $Q^2$ at fixed $x_B$ and $t$ the
scattering amplitude factorizes into a hard-scattering kernel,
generalized quark or gluon distributions, and the light-cone
distribution amplitude of the produced meson \cite{Collins:1996fb}.
We make the approximation that the normalization of the $\rho$ and
$\phi$ distribution amplitudes is related by $\langle \rho| \bar{u}
\gamma^\mu u - \bar{d} \gamma^\mu d|0\rangle = \sqrt{2} \langle \phi|
\bar{s} \gamma^\mu s|0\rangle$.  This relation leads to a value of
$9:2$ for the ratio $(M_\rho \Gamma_{\rho\to e^+e^-}) : (M_\phi
\Gamma_{\phi\to e^+e^-})$ of meson mass times partial leptonic width,
which compares well with the value $9: 2.1$ from experiment
\cite{Eidelman:2004wy}.  We further assume that the $\rho$ and $\phi$
distribution amplitudes have the same dependence on the quark momentum
fraction.  The ratio of production amplitudes for the two channels is
then\footnote{We remark that there is a mistake in eq.~(284) of
\protect\cite{die:03}: in all three terms with $F^g$ the 8 in the
prefactor should be replaced by 4.}
\begin{equation}
  \label{amp-ratio}
\mathcal{A}_{\rho} : \mathcal{A}_{\phi} =
-\frac{1}{\sqrt{2}} \Bigg( \frac{2}{3}\, \mathcal{F}^u 
  + \frac{1}{3}\, \mathcal{F}^d + \frac{3}{4}\, \mathcal{F}^g \Bigg) :
\Bigg( \frac{1}{3}\, \mathcal{F}^s + \frac{1}{4}\, \mathcal{F}^g \Bigg) 
\end{equation}
to leading accuracy in $1/Q$ and in $\alpha_S$.  Here
\begin{eqnarray}
  \label{form-factors}
\mathcal{F}^q &=& \int_0^1 dx\, \Bigg[
  \frac{1}{\xi-x - i\varepsilon} - \frac{1}{\xi+x - i\varepsilon} \Bigg]
   \Big[ F^q(x,\xi,t) - F^q(-x,\xi,t) \Big]
\quad\quad (q=u,d,s) \, ,
\nonumber \\
\mathcal{F}^g &=& \int_0^1 dx\, \Bigg[
  \frac{1}{\xi-x - i\varepsilon} - \frac{1}{\xi+x - i\varepsilon} \Bigg]
   \frac{F^g(x,\xi,t)}{x}
\end{eqnarray}
with $\xi= x_B/(2-x_B)$ are the relevant integrals over quark and
gluon matrix elements, para\-meterized by generalized parton
distributions as
\begin{equation}
F^q(x,\xi,t) = \frac{1}{(p+p')^+} \left[
  H^q(x,\xi,t)\, \bar{u}(p') \gamma^+ u(p) +
  E^q(x,\xi,t)\, \bar{u}(p') 
                 \frac{i \sigma^{+\mu} (p'-p)_\mu}{2m}\, u(p)
  \, \right]
\end{equation}
for quarks and in analogy for gluons.  The distributions are
normalized such that in the forward limit and for $x>0$ one has
$H^q(x,0,0) = q(x)$, $H^q(-x,0,0) = - \bar{q}(x)$ and $H^g(x,0,0) = x
g(x)$, for explicit definitions see e.g.\ \cite{die:03}.  It is
understood that the distributions are to be taken at a factorization
scale of order $Q^2$.  We restrict our study to the Born level
formulae (\ref{amp-ratio}) and (\ref{form-factors}) and note that at
next-to-leading order in $\alpha_S$ the amplitudes depend in addition
on the quark flavor singlet distribution $\sum_q [ F^q(x,\xi,t) -
F^q(-x,\xi,t)]$, which mixes with $F^g(x,\xi,t)$ under
evolution~\cite{Ivanov:2004zv}.

%%%%%%%%%%%%%%%%%%%%%%%%%%%%%%%%%%%%%%%%%%%%
\vskip \baselineskip
\noindent 3.\,
The $\gamma^* p$ cross section on an unpolarized target involves the
combination
\begin{eqnarray}
  \label{unpol-xsect}
\frac{1}{2} \sum_{s's} 	|\mathcal{F}_{s's}|^2
  &=& (1-\xi^2)\, |\mathcal{H}|^2
   - \left( \xi^2 + \frac{t}{4m^2} \right) |\mathcal{E}|^2 
   - 2 \xi^2 \re (\mathcal{E}^* \mathcal{H} ) \, ,
\end{eqnarray}
where $s$ and $s'$ respectively denote the polarization of the initial
and final state proton, and where $\mathcal{F}$, $\mathcal{H}$ and
$\mathcal{E}$ are the relevant linear combinations of integrals over
quark and gluon distributions given in (\ref{amp-ratio}).  In the
following we will be interested in kinematics where $\xi$ is below 0.1
and where the dominant values of $-t$ are a few times $0.1 \gev^2$.
Both $\xi^2$ and $t/(4 m^2)$ are then small, so that the term with
$|\mathcal{H}|^2$ will dominate the unpolarized cross section unless
$\mathcal{E}$ is significantly larger than $\mathcal{H}$.

There are however no indications from theory or phenomenology that
$|E^q| \gg |H^q|$ or $|E^g| \gg |H^g|$.  For their lowest $x$ moments,
one has for instance
\begin{equation}
\textstyle
\int_{-1}^1 dx\, H^u = 2 , \quad
\int_{-1}^1 dx\, H^d = 1 , \quad
\int_{-1}^1 dx\, E^u \approx 1.67 , \quad
\int_{-1}^1 dx\, E^d \approx -2.03 ,
\end{equation}
at $t=0$, where the integrals over $E^q$ are obtained from the
anomalous magnetic moments of proton and neutron.  A similar situation
is seen when comparing the moments $\int_{-1}^1 dx\, x H^q$ and
$\int_{-1}^1 dx\, x E^q$ for $u$ and $d$ quarks obtained in lattice
QCD \cite{goe:03,hae:03}.  For the gluon distributions, one may even
expect that $E^g$ is relatively small.  For the following argument we
set $\xi=0$ and $t=0$.  We then have a sum rule 
\begin{equation}
  \label{sum-rule}
\textstyle
\int_{0}^1 dx\, E^g + \sum_q \int_{-1}^1 dx\, x E^q = 0
\end{equation}
from the conservation of momentum and angular momentum (see e.g.\
\cite{die:03}).  The lattice calculations just cited find that the
contributions from $u$ and $d$ quarks tend to cancel each other, in
line with general considerations from the large $N_c$ limit of QCD
\cite{goe:01}.  Depending on how strong this cancellation is, and
barring an unexpectedly large contribution from $s$ quarks to the sum
rule (\ref{sum-rule}), the integral $\int_{0}^1 dx\, E^g$ can thus be
relatively small compared with the individual quark moments
$\int_{-1}^1 dx\, x E^u$ and $\int_{-1}^1 dx\, x E^d$.  This is in
stark contrast to the momentum sum $\int_{0}^1 dx\, H^g = \int_{0}^1
dx\, x g(x)$, which is similar in size to its quark counterparts.

In the following we will thus neglect $\mathcal{E}$ in the unpolarized
cross section (\ref{unpol-xsect}) because of its kinematic prefactors.
We note in passing that the difference of $\gamma^* p$ cross sections
for transverse target polarization above and below the scattering
plane is proportional to $\im ( \mathcal{E}^* \mathcal{H} )$, so that
in this observable the contribution of $\mathcal{E}$ is essential.  We
observe that with the pattern of relative signs and sizes just
discussed, the contributions from $E^u$ and $E^d$ in $\rho$ production
will partially cancel according to (\ref{amp-ratio}).

\begin{figure}
\begin{center}
\includegraphics[width=0.49\textwidth]{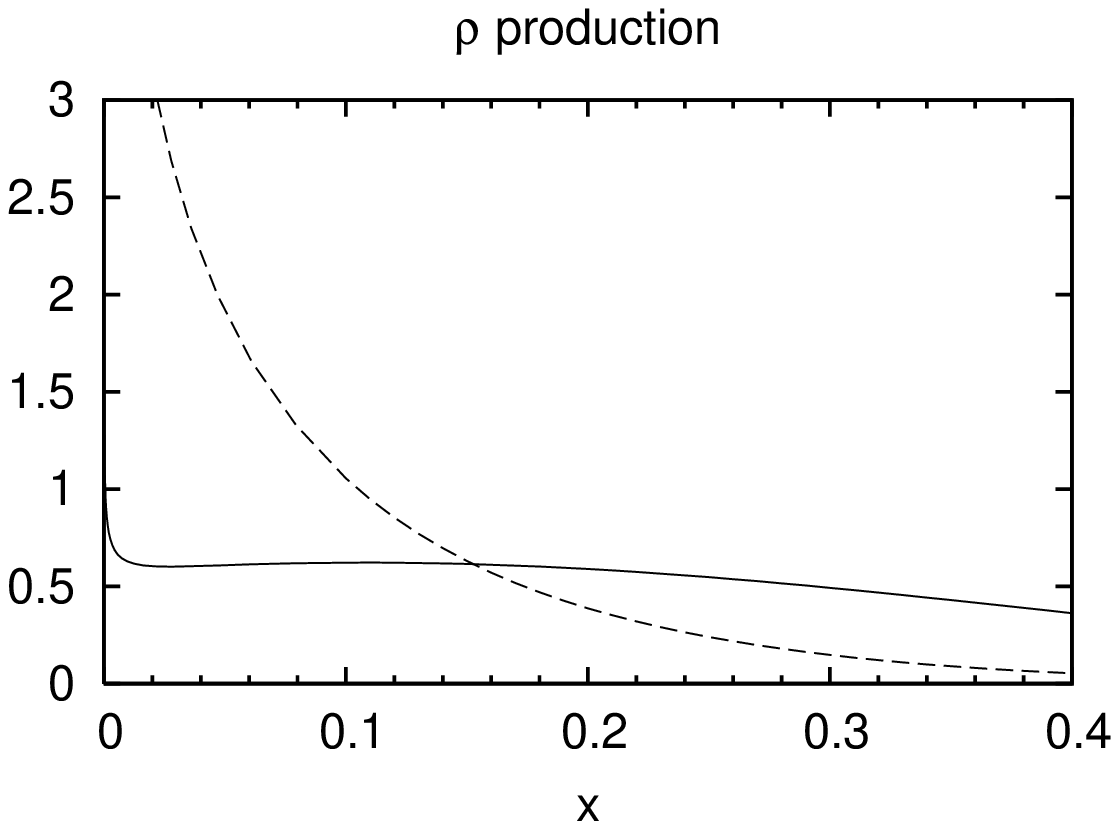}
\includegraphics[width=0.49\textwidth]{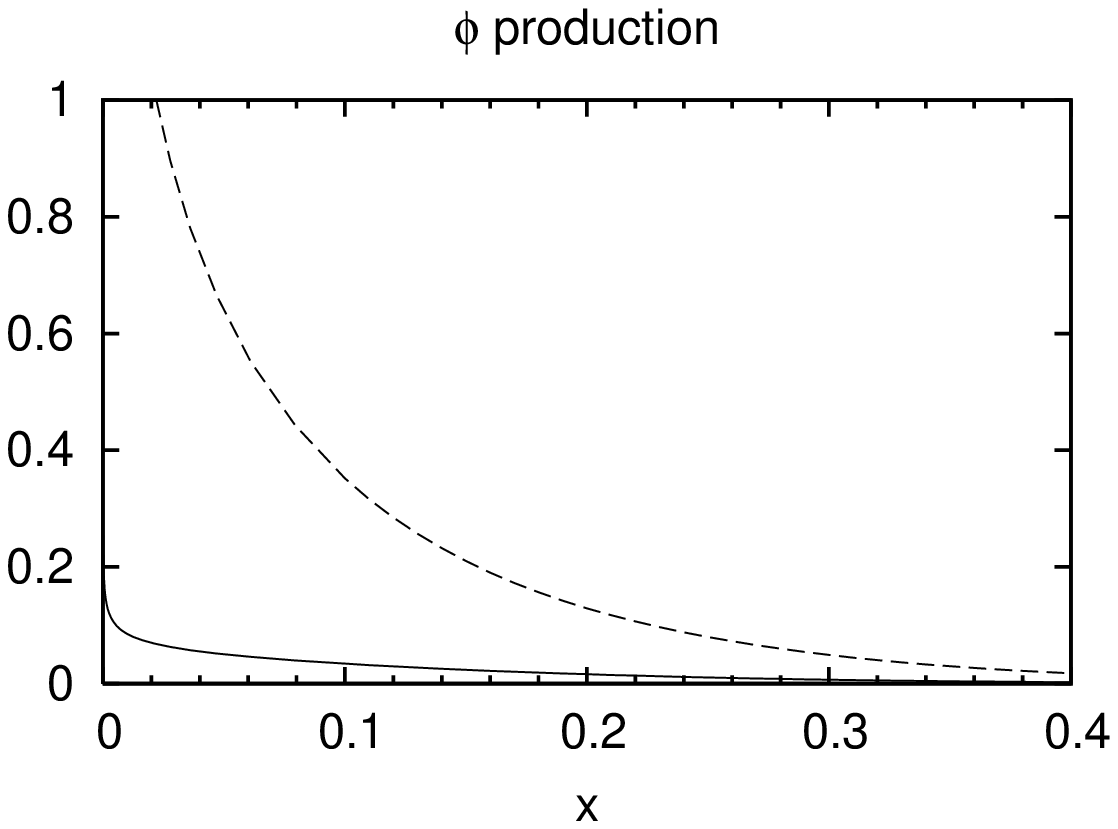}
\caption{\label{fig:forward} Left: The combinations $\frac{2}{3}
  (u+\bar{u}) + \frac{1}{3} (d+\bar{d})$ and $\frac{3}{4} g$ of parton
  distributions relevant for $\rho$ production according to
  (\protect\ref{amp-ratio}) and (\protect\ref{form-factors}).  Right:
  The combinations $\frac{1}{3} (s+\bar{s})$ and $\frac{1}{4} g$
  relevant for $\phi$ production.  Full lines correspond to quarks and
  dashed lines to gluons.}
\end{center}
\end{figure}

%%%%%%%%%%%%%%%%%%%%%%%%%%%%%%%%%%%%%%%%%%%%
\vskip \baselineskip
\noindent 4.\,
For a very rough estimate of the relative importance of the different
terms in (\ref{amp-ratio}) let us neglect that the dependence of
generalized parton distributions on $\xi$ and on $t$ is most likely
not the same for quarks and for gluons.  Taking $\xi=0$ and $t=0$ one
is then led to compare the combinations of parton densities shown in
Fig.~\ref{fig:forward}, keeping in mind that the dominant values of
$x$ in the convolutions (\ref{form-factors}) are generically of order
$\xi$.  We show the CTEQ6L distributions \cite{CTEQ} at a scale of
$\mu=2 \gev$ and remark that the situation is qualitatively similar
for the starting scale $\mu=1.3 \gev$ of the CTEQ6L parameterization.

The dominance of gluons over sea quarks in the case of $\phi$
production is so pronounced that, despite the simplicity of our
estimate, we will in the following neglect $\mathcal{F}^s$.  For the
$\rho$ channel the estimate suggests that gluons become important
already at moderately small $x_B$.  This expectation will now be
confronted with data.

%%%%%%%%%%%%%%%%%%%%%%%%%%%%%%%%%%%%%%%%%%%%
\vskip \baselineskip
\noindent 5.\,
With the approximations discussed so far, the ratio of $\phi$ and
$\rho$ production cross sections is
\begin{equation}
  \label{xsect-ratio}
\frac{\sigma_{\phi}}{\sigma_{\rho}} \approx
  \frac{2}{9}\, \frac{|g_\rho|^2}{|g_\rho|^2 + 2 |g_\rho| |q_\rho|
  \cos\alpha + |q_\rho|^2}
\end{equation}
with
\begin{eqnarray}
&& |g_\rho|^2 
   = \int\! dt\, \Big| \frac{3}{4}\, \mathcal{H}^g\, \Big|^2 \, , 
\qquad\qquad
|q_\rho|^2 
   = \int\! dt\, \Big| \frac{2}{3}\, \mathcal{H}^u 
                   + \frac{1}{3}\, \mathcal{H}^d\, \Big|^2 \, ,
\nonumber \\
&& |g_\rho| |q_\rho| \cos\alpha 
   = \re \int\! dt\, \frac{3}{4}\, \mathcal{H}^g \,
   \Big( \frac{2}{3}\, \mathcal{H}^u + \frac{1}{3}\, \mathcal{H}^d
   \Big)^* \, .
\end{eqnarray}
$\alpha$ may be regarded as the ``average phase'' between gluon and
quark amplitudes, where the ``average'' is over $t$.

\begin{figure}
\begin{center}
\includegraphics[width=0.55\textwidth]{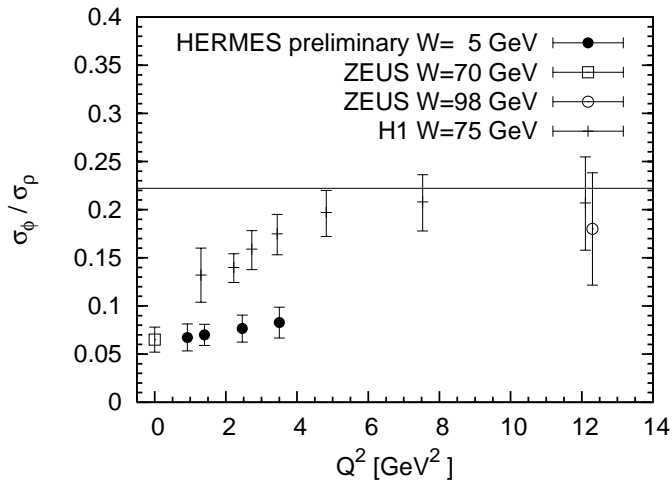}
\caption{\label{fig:data} Data for the cross section ratio
$\sigma_{\phi} /\sigma_{\rho}$.  Statistical and systematic errors
given in \protect\cite{Derrick:1996nb,Adloff:2000nx} are added in
quadrature.  The horizontal line indicates the value $2/9$
corresponding to a negligible quark exchange contribution in both
$\phi$ and $\rho$ production.}
\end{center}
\end{figure}

Preliminary data from HERMES \cite{Borissov:2001fq} on $\sigma_{\phi}
/\sigma_{\rho}$ at are shown in Fig.~\ref{fig:data}, together with
results at very small $x_B$ from ZEUS and H1
\cite{Derrick:1996af,Derrick:1996nb,Adloff:2000nx}.  The two HERMES
points with $Q^2=2.46\gev^2$ and $3.5 \gev^2$ respectively correspond
to $x_B=0.09$ and $x_B=0.13$, with cross section ratios of $0.0765\pm
0.014$ and $0.0827\pm 0.016$.  In the following we will take
$\sigma_{\phi}/\sigma_{\rho} = 0.08$ for simplicity.  Inverting
(\ref{xsect-ratio}) one obtains
\begin{equation}
|\, q_\rho /g_\rho | = -\cos\alpha
 + \sqrt{ (2\sigma_\rho)/(9\sigma_\phi) - \sin^2\alpha } \, ,
\end{equation}
where we used that the solution with a minus instead of a plus in
front of the square root is not admissible for
$\sigma_{\phi}/\sigma_{\rho} <2/9$.  This gives
\begin{equation}
  \label{master-estimate}
  0.38 \le |\, g_\rho / q_\rho | \le 1.5 \, ,
\end{equation}
with the values $0.38$ and $1.5$ respectively corresponding to
$\cos\alpha=-1$ and $\cos\alpha=1$.  We remark that with current
models of generalized parton distributions one typically finds that
the convolution integrals (\ref{form-factors}) tend to be dominated by
their imaginary parts and that these are positive, so that one may
regard values of $\cos\alpha$ near 1 as more likely.  Our result
(\ref{master-estimate}) thus agrees rather well with the simple
estimate one can obtain from Fig.~\ref{fig:forward}.

%%%%%%%%%%%%%%%%%%%%%%%%%%%%%%%%%%%%%%%%%%%%
\vskip \baselineskip
\noindent 6.\,
It is well known that for vector meson production at $Q^2$ of a few
$\gev^2$ there are sizeable power corrections (see
\cite{goe:01,die:03} and references therein).  In particular, the
leading approximation in $1/Q$ cannot account for the normalization of
the meson production cross section.  Some but not all of the power
corrections will partially cancel in the cross section ratio for
$\phi$ and $\rho$ production.

The H1 and ZEUS electroproduction data in Fig.~\ref{fig:data} show a
clear increase of $\sigma_\phi /\sigma_\rho$ with $Q^2$ in a region of
$x_B$ where one would expect both $\phi$ and $\rho$ production to be
clearly dominated by gluon exchange.  It is tempting to ascribe this
to an additional suppression of $\sigma_\phi$ compared to
$\sigma_\rho$ at moderate $Q^2$ due to effects of the strange quark
mass in the quark loop of the hard scattering graphs (see
Fig.~\ref{fig:mesons}).  There is another effect contributing to this
trend which already appears at leading order in $1/Q$ and $\alpha_S$.
In this approximation the cross section is proportional to the square
of the integral $\int_0^1 dz\, z^{-1} (1-z)^{-1}\, \phi(z)$ over the
meson distribution amplitude, where $z$ is the momentum fraction of
the quark.  The above trend will thus be enhanced if at lower
factorization scales $Q^2$ the distribution amplitude is narrower for
the $\phi$ than for the $\rho$ and only evolves to a similar shape for
the two mesons with increasing $Q^2$.

If a similar suppression of $\sigma_\phi$ compared with $\sigma_\rho$
also takes place in the kinematics of the HERMES measurement, then the
right-hand-side of (\ref{xsect-ratio}) must be multiplied with a
correction factor below 1, leading to an even larger estimate for $|\,
g_\rho / q_\rho |$ than in (\ref{master-estimate}).

%%%%%%%%%%%%%%%%%%%%%%%%%%%%%%%%%%%%%%%%%%%%
\vskip \baselineskip
\noindent 7.\,
The leading approximation in $1/Q$ predicts longitudinal polarization
of both the virtual photon and the produced vector meson
\cite{Collins:1996fb}.  Experimentally one finds that both in $\rho$
and in $\phi$ production the ratio $R = \sigma_L /\sigma_T$ of cross
sections for longitudinal and transverse photon polarization is not
very large for $Q^2$ of a few $\gev^2$.  This is another example that
$1/Q$ suppressed effects are not entirely negligible in this kinematic
region.  In the relation (\ref{xsect-ratio}) we should more precisely
have the ratio $\sigma_{L \phi} /\sigma_{L \rho}$ of longitudinal
cross sections on the left-hand-side.  Preliminary HERMES data
\cite{Rakness:PhD,Tytgat:PhD,Borissov:2001jj} suggest that $R_\phi$
may be somewhat smaller than $R_\rho$ in the kinematics of the HERMES
points in Fig.~\ref{fig:data}.  This would correspond to $\sigma_{L
\phi} /\sigma_{L \rho} < \sigma_{\phi} /\sigma_{\rho}$ and again lead
to a larger estimate for $|\, g_\rho / q_\rho |$ than in
(\ref{master-estimate}).

%%%%%%%%%%%%%%%%%%%%%%%%%%%%%%%%%%%%%%%%%%%%
\vskip \baselineskip
\noindent 8.\,
In summary, we find that under rather weak assumptions the HERMES data
on the ratio of $\phi$ and $\rho$ electroproduction cross sections
indicates a substantial contribution from gluon exchange in the $\rho$
channel for $x_B$ around 0.1 and $Q^2$ of a few $\gev^2$.

This conclusion is in contrast with the results of
\cite{Vanderhaeghen:1999xj}, shown for the kinematics of the HERMES
measurement in \cite{air:00}, where the gluon contribution to the
cross section was estimated to be quite small.  We cannot fully
resolve this discrepancy here, but remark that for the gluon exchange
contribution to $\rho$ production Ref.~\cite{Vanderhaeghen:1999xj}
used the calculation of \cite{fra:96}, which was performed for the
limit of very small $x_B$.  It is difficult to assess the reliability
of extrapolating the results of \cite{fra:96} to $x_B \sim 0.1$.  A
model calculation of both quark and gluon contributions in $\rho$
production based on the convolutions in (\ref{form-factors}) is under
way \cite{ell:04}.

%%%%%%%%%%%%%%%%%%%%%%%%%%%%%%%%%%%%%%%%%%%%
\vskip \baselineskip
\noindent \textbf{Acknowledgments}.\,
We thank E.-C.~Aschenauer, A.~Borissov and W.-D.~Nowak for valuable
discussions or correspondence.  A.V.\ is supported by RFBR grants
04-02-16445 and 03-02-17291 and by the Heisenberg-Landau program.

%%%%%%%%%%%%%%%%%%%%%%%%%%%%%%%%%%%%%%%%%%%%


\begin{thebibliography}{99}

\bibitem{mul:94}
D.~M{\"u}ller, D.~Robaschik, B.~Geyer, F.~M.~Dittes and J.~Ho\v{r}ej\v{s}i,
%``Wave functions, evolution equations and evolution kernels from light-ray
%operators of {QCD},''
Fortsch.\ Phys.\ {\bf 42}, 101 (1994)
[hep-ph/9812448]; \\
%%CITATION = HEP-PH 9812448;%%
%
%\bibitem{rad:97}
A.~V.~Radyushkin,
%``Nonforward parton distributions,''
Phys.\ Rev.\ D {\bf 56}, 5524 (1997)
[hep-ph/9704207]; \\
%%CITATION = HEP-PH 9704207;%%
%
%\bibitem{Blumlein:1997pi}
J.~Bl{\"u}mlein, B.~Geyer and D.~Robaschik,
 %``On the evolution kernels of twist 2 light-ray operators for
 %unpolarized  and polarized deep inelastic scattering,''
Phys.\ Lett.\ B {\bf 406}, 161 (1997)
[hep-ph/9705264].
%%CITATION = HEP-PH 9705264;%%

\bibitem{ji:96}
X.~D.~Ji,
Phys. Rev. Lett. {\bf 78}, 610 (1997) 
[hep-ph/9603249].
%%CITATION = HEP-PH 9603249;%%

\bibitem{ral:02}
J.~P. Ralston and B.~Pire,
Phys. Rev. {\bf D66}, 111501 (2002)
[hep-ph/0110075].
%%CITATION = HEP-PH 0110075;%%

\bibitem{bur:02} 
M.~Burkardt,
%``Impact parameter space interpretation for generalized parton
%distributions,''
Int.\ J.\ Mod.\ Phys.\ A {\bf 18}, 173 (2003)
[hep-ph/0207047];\\
%%CITATION = HEP-PH 0207047;%%
%
%\bibitem{Diehl:2002he}
M.~Diehl,
%``Generalized parton distributions in impact parameter space,''
Eur.\ Phys.\ J.\ C {\bf 25}, 223 (2002)
[hep-ph/0205208].
%%CITATION = HEP-PH 0205208;%%

\bibitem{goe:01}
K.~Goeke, M.~V.~Polyakov and M.~Vanderhaeghen,
%``Hard exclusive reactions and the structure of hadrons,''
Prog.\ Part.\ Nucl.\ Phys.\  {\bf 47}, 401 (2001)
[hep-ph/0106012].
%%CITATION = HEP-PH 0106012;%% 

\bibitem{Belitsky:2001ns}
A.~V.~Belitsky, D.~M\"{u}ller and A.~Kirchner,
%``Theory of deeply virtual Compton scattering on the nucleon,''
Nucl.\ Phys.\ B {\bf 629}, 323 (2002)
[hep-ph/0112108].
%%CITATION = HEP-PH 0112108;%%

\bibitem{die:03} 
M.~Diehl,
%``Generalized parton distributions,''
Phys.\ Rept.\  {\bf 388}, 41 (2003)
[hep-ph/0307382].
%%CITATION = HEP-PH 0307382;%%

\bibitem{hermes:00}
A.~Airapetian {\it et al.}\ [HERMES Collaboration],
%``Exclusive leptoproduction of rho0 mesons from hydrogen at intermediate
%virtual photon energies,''
Eur.\ Phys.\ J.\ C {\bf 17}, 389 (2000)
[hep-ex/0004023].
%%CITATION = HEP-EX 0004023;%%

\bibitem{Collins:1996fb}
J.~C.~Collins, L.~Frankfurt and M.~Strikman,
%``Factorization for hard exclusive electroproduction of mesons in
%QCD,'' 
Phys.\ Rev.\ D {\bf 56}, 2982 (1997)
[hep-ph/9611433].
%%CITATION = HEP-PH 9611433;%%

\bibitem{Eidelman:2004wy}
S.~Eidelman {\it et al.}  [Particle Data Group],
%``Review of particle physics,''
Phys.\ Lett.\ B {\bf 592}, 1 (2004).
%%CITATION = PHLTA,B592,1;%%

\bibitem{Ivanov:2004zv}
D.~Y.~Ivanov, L.~Szymanowski and G.~Krasnikov,
%``Vector meson electroproduction at next-to-leading order,''
JETP Lett.\  {\bf 80}, 226 (2004)
[Pisma Zh.\ Eksp.\ Teor.\ Fiz.\  {\bf 80}, 255 (2004)]
[hep-ph/0407207].
%%CITATION = HEP-PH 0407207;%%

\bibitem{goe:03} 
M.~G{\"o}ckeler {\it et al.}\
% R.~Horsley, D.~Pleiter, P.~E.~L.~Rakow, A.~Sch{\"a}fer, G.~Schierholz and
% W.~Schroers 
[QCDSF Collaboration],
%``Generalized parton distributions from lattice QCD,''
Phys.\ Rev.\ Lett.\  {\bf 92}, 042002 (2004)
[hep-ph/0304249].
%%CITATION = HEP-PH 0304249;%%

\bibitem{hae:03}
P.~H{\"a}gler {\it et al.}\
% J.~W.~Negele, D.~B.~Renner, W.~Schroers, T.~Lippert and K.~Schilling 
[LHPC and SESAM Collaborations],
Phys.\ Rev.\ D {\bf 68}, 034505 (2003)
hep-lat/0304018.
%%CITATION = HEP-LAT 0304018;%%

\bibitem{CTEQ} 
J. Pumplin {\it et al.}\ [CTEQ collaboration],
%``New generation of parton distributions with uncertainties from
%global QCD analysis,''
JHEP {\bf 0207}, 012 (2002)
[hep-ph/0201195].
%%CITATION = HEP-PH 0201195;%%

\bibitem{Borissov:2001fq}
A.~B.~Borissov  [HERMES Collaboration],
%``Diffraction and deeply virtual exclusive scattering at HERMES,''
Nucl.\ Phys.\ Proc.\ Suppl.\  {\bf 99A}, 156 (2001).
%%CITATION = NUPHZ,99A,156;%%

\bibitem{Derrick:1996af}
M.~Derrick {\it et al.}\  [ZEUS Collaboration],
%``Measurement of Elastic $\phi$ Photoproduction at HERA,''
Phys.\ Lett.\ B {\bf 377}, 259 (1996)
[hep-ex/9601009].
%%CITATION = HEP-EX 9601009;%%

\bibitem{Derrick:1996nb}
M.~Derrick {\it et al.}\ [ZEUS Collaboration],
%``Measurement of the Reaction $\gamma~* p \to \phi p$ in Deep
%Inelastic $e~+p$ Scattering at HERA,''
Phys.\ Lett.\ B {\bf 380}, 220 (1996)
[hep-ex/9604008].
%%CITATION = HEP-EX 9604008;%%

\bibitem{Adloff:2000nx}
C.~Adloff {\it et al.}\  [H1 Collaboration],
%``Measurement of elastic electroproduction of Phi mesons at HERA,''
Phys.\ Lett.\ B {\bf 483}, 360 (2000)
[hep-ex/0005010].
%%CITATION = HEP-EX 0005010;%%

\bibitem{Rakness:PhD}
G.~L.~Rakness, Doctoral Thesis, University of Colorado at Boulder,
2000. 

\bibitem{Tytgat:PhD}
M.~Tytgat, Doctoral Thesis, Universiteit Gent, 2000.

\bibitem{Borissov:2001jj}
A.~B.~Borissov  [HERMES Collaboration],
%``Spin physics of vector meson production at intermediate W and Q**2,''
Procs.\ of the 9th International Workshop on High-Energy Spin Physics
(SPIN 01), Dubna, Russia, 2--7 Aug.~2001,
DESY-HERMES-01-60.

\bibitem{Vanderhaeghen:1999xj}
M.~Vanderhaeghen, P.~A.~M.~Guichon and M.~Guidal,
%``Deeply virtual electroproduction of photons and mesons on the nucleon:
%Leading order amplitudes and power corrections,''
Phys.\ Rev.\ D {\bf 60}, 094017 (1999)
[hep-ph/9905372].
%%CITATION = HEP-PH 9905372;%%

\bibitem{air:00}
A.~Airapetian {\it et al.}\ [HERMES Collaboration],
%``Exclusive leptoproduction of rho0 mesons from hydrogen at intermediate
%virtual photon energies,''
Eur.\ Phys.\ J.\ C {\bf 17}, 389 (2000)
[hep-ex/0004023].
%%CITATION = HEP-EX 0004023;%%

\bibitem{fra:96}
L.~Frankfurt, W.~Koepf and M.~Strikman,
%``Hard diffractive electroproduction of vector mesons in QCD,''
Phys.\ Rev.\ D {\bf 54}, 3194 (1996)
[hep-ph/9509311].
%%CITATION = HEP-PH 9509311;%%

\bibitem{ell:04}
F.~Ellinghaus, W.-D.~Nowak, A.~V.~Vinnikov and Z.~Ye,
in preparation.

\end{thebibliography}
\end{document}